\documentstyle[preprint,revtex]{aps}
\begin{document}
\draft
\begin{title}
\vskip4.0cm
{Ghost Poles in the Nucleon Propagator:}\\
{Vertex Corrections and Form Factors}
\end{title}
\author{G. Krein}
\begin{instit}
{Instituto de F\'{\i}sica Te\'orica - Universidade Estadual Paulista}\\
{Rua Pamplona, 145 - 01405 S\~ao Paulo - Brazil}
\end{instit}
\moreauthors{M. Nielsen}
\begin{instit}
{Instituto de F\'{\i}sica - Universidade de S\~ao Paulo}\\
{Caixa Postal, 20516 - 01498 S\~ao Paulo - Brazil}
\end{instit}
\moreauthors{R.D. Puff and L. Wilets}
\begin{instit}
{Department of Physics FM-15, University of Washington,
Seattle WA 98195 - USA}
\end{instit}
\newpage
\begin{abstract}
\vskip3.0cm
\centerline{Abstract}
\vskip1.0cm
Vertex corrections are taken into account in the Schwinger-Dyson equation
for the nucleon propagator in a relativistic field theory of fermions and
mesons. The usual Hartree-Fock approximation for the nucleon propagator is
known to produce the appearance of complex (ghost) poles which violate basic
theorems of quantum field theory. In a theory with vector mesons there are
vertex corrections that produce a strongly damped vertex function in the
ultraviolet. One set of such corrections is known as the Sudakov form factor
in quantum electrodynamics. When the Sudakov form factor generated by massive
neutral vector mesons is included in the Hartree-Fock approximation to the
Schwinger-Dyson equation for the nucleon propagator, the ghost poles disappear
and consistency with basic requirements of quantum field theory is recovered.
\end{abstract}
\vskip1.0cm
\newpage
\begin{narrowtext}
\section{INTRODUCTION}
Non-relativistic many-body theory has been used with success in the study
of ground and excited states of nuclear systems. Nevertheless,
experiments in the near future will provide data on nuclear systems at
extreme conditions of density and temperature and obviously a new theoretical
approach that goes beyond the non-relativistic one will be required. This
has motivated a great deal of interest in recent years in the development
of relativistic many-body theories for nuclear physics based on the methods
of renormalizable relativistic quantum field theory. In this context, there is
an extensive literature on calculations of nuclear matter and finite nuclei
properties by means of models based on the original Walecka model\cite{wa}.
Calculations employing mean field and one loop (Hartree) approximations have
achieved considerable success in the description of bulk properties of nuclei
and of proton-nucleus scattering parameters. However, there are severe
difficulties in extending the calculations to include quantum corrections
which go beyond the one-loop Hartree approximation. The inclusion of these
quantum corrections leads to catastrophic results, with the appearance of
complex poles in the baryon and meson propagators which, among other things,
introduce a large imaginary part to the nuclear matter energy.

Complex poles, or ghosts, have long been noted in local relativistic field
theory \cite{{all},{bpw}}.
They are physically unacceptable because they correspond to eigenstates
of the system with complex energies and probabilities. In the case of quantum
electrodynamics(QED), the appearance of a ghost in the one loop correction to
the photon propagator, the so called Landau ghost, is not taken as a serious
drawback of the theory. This is because the momentum scale at which the ghost
appears is far from measurable and, at this scale QED should probably be
modified to include the effects of other electroweak effects. In the case of a
nuclear theory with mesons and baryons, the corresponding ghosts appear at
momentum scales of the order of one $GeV$. The appearance of the ghost poles
is related to the short distance behavior of the model interactions;
asymptotically free theories appear to be free of ghosts\cite{pe}. It is clear
that a description of hadronic matter in terms of mesons and baryons only must
break down in the region where short distance properties are involved.
However, such a description should provide a reasonable description of the
properties which are thought to be insensitive to the short distance physics.
In this sense, the appearance of ghosts in hadronic theories appears to
frustrate the hope of constructing relativistic models based on renormalizable
field theories without employing subnucleonic degrees of freedom.
Achtzehnter and Wilets\cite{aw} have shown that quark sub-structure plays an
important role in the interaction of nucleons with an external (Bose) field
{\it at all momentum transfers}.  This involves issues which we will not
pursue here.  Rather, we concentrate on the construction of some intrinsically
consistent field theory. Of course
one should keep in mind that the effective Lagrangians commonly used in nuclear
physics are likely not be derivable from the fundamental theory of the
strong interactions (QCD). Such an effective Lagrangian will probably be very
complicated and quite inelegant.

In the past, several methods have been proposed to
eliminate this short-distance sensitivity as, for example, modifying the
analytic structure of the propagators in such a way to remove the unwanted
singularities\cite{re}. More recently, in the light of the quark substructure
of the nucleons, form factors at the meson-baryon vertices\cite{pra}
have been used to regulate the theory at short distances. Another perspective
on the problem is the regulation of the theory by means of vector meson
dressing of nucleon-meson vertices. It is known that in a theory with neutral
vector mesons there are vertex corrections that generate a strongly damped
vertex function in the ultraviolet\cite{known}. The damping arises from the
infrared structure of the theory, despite the fact that the external nucleon
momenta and the momentum transfer to the vertex are large. Physically, the
damping arises from the large likelihood of matter fields to radiate soft
virtual vector mesons. This phenomenon is a property of theories containing
vector mesons. Based on these considerations in QED, Sudakov\cite{sud}
derived a form factor. In a recent publication, Allendes and Serot\cite{alse}
included the Sudakov form factor in the calculation of the
polarization loop correction to the vector meson propagator. The authors
concluded that inclusion of the Sudakov form factor regulates the ultraviolet
behavior so that the corrected propagator is free from ghost poles. This is
a very important achievement for the construction of a consistent relativistic
nuclear many-body theory with mesons and baryons only, as it restores the
hope for the consistency of such a theory.

In this paper we study the effect of the Sudakov and quark-substructure form
factors on the ghost problem in the nucleon propagator. We consider a model
with nucleons, neutral vector mesons and pions. Although the theory is designed
to study nuclear matter, we shall restrict ourselves in this paper for
simplicity to the vacuum only. The appearance of the ghosts does not depend
on the presence of matter\cite{wpcn}. We analyze the effect of the
vertex corrections by means of form factors in the renormalized
Schwinger-Dyson equation for the nucleon propagator. The calculation of the
complete vertex function is a complicated problem and therefore we shall
use simplifying approximations which we discuss below.

\section{NUCLEON PROPAGATOR AND GHOST POLES}
In order to make the paper self-contained, at the cost of being a little
repetitive, in the following we briefly review the problem of ghosts in the
nucleon propagator following the work of Brown, Puff and Wilets(BPW)\cite{bpw}.
We start with the usual definition of the nucleon propagator

\begin{equation}
G_{\beta \alpha}(x-x')=-i<0|T[\psi_{\beta}(x')\bar \psi_{\alpha}(x)]|0>\;,
\label{defpr}
\end{equation}

\noindent
where $\psi$ represents the nucleon field operator and $|0>$ is the physical
vacuum state. The K\"allen-Lehmann representation for the Fourier transform
$G(p)$ of $G_{\beta \alpha}(x-x')$ can be written as

\begin{equation}
G(p)=\int_{- \infty}^{+ \infty} d\kappa {A(\kappa) \over {{\not\!p} - \kappa
+ i\epsilon}}\;.
\label{kaleh}
\end{equation}

\noindent
$A(\kappa)$ is the spectral function. It represents the probability that a
state of mass $|\kappa|$ is created by $\psi$ or $\bar \psi$, and as such it
must be non-negative. Negative $\kappa$ corresponds to states with
opposite parity to the nucleon.

Eq. (\ref{kaleh}) can be rewritten conveniently as

\begin{equation}
G(p)=P_{+}(p)\tilde G (w_p+i\epsilon)+P_{-}(p)\tilde G(-w_p-i\epsilon)\;,
\label{gpro}
\end{equation}

\noindent
where $P_{\pm}(p)$ are projection operators defined as

\begin{equation}
P_{\pm}(p)={1\over 2}\left(1 \pm {{\not\!p} \over w_p}\right)\;,
\end{equation}

\noindent
with

\begin{eqnarray}
w_p = \sqrt{p^2} =
          \left\{ \begin{array}{ll}
                     \sqrt{p^2}, & \mbox{if $p^2 > 0$}; \\
                    i\sqrt{-p^2}, & \mbox{if $p^2 < 0$.},
                              \end{array}
                              \right.\;,
\label{wp}
\end{eqnarray}

\noindent
and $\tilde G(z)$ is given by the dispersion integral

\begin{equation}
\tilde G(z) = \int_{- \infty}^{+ \infty} d\kappa
{A(\kappa) \over {z - \kappa}}\;.
\label{gtil}
\end{equation}

\noindent
It follows from the commutation relations that

\begin{equation}
\int_{- \infty}^{+ \infty} d\kappa A(\kappa) = 1\;.
\label{norm}
\end{equation}

\noindent
The inverse of the propagator can be written in terms of the projection
operators $P_{\pm}(p)$ as

\begin{equation}
G^{-1}(p)=P_{+}(p)\tilde G^{-1} (w_p+i\epsilon)+
P_{-}(p)\tilde G^{-1}(-w_p-i\epsilon)\;.
\label{ginvpro}
\end{equation}

Since $A(\kappa)$ is supposed to be non-negative, it is simple to show that
$\tilde G(z)$ can have no poles or zeros off the real axis. This is
known as the Herglotz property. Now, if $\tilde G(z)$ possesses the Herglotz
property, then so does $\tilde G^{-1}(z)$. This permits us to write a
spectral representation for $\tilde G^{-1}(z)$,

\begin{equation}
\tilde G^{-1}(z)=z-M_0 - \int_{- \infty}^{+ \infty} d\kappa
{T(\kappa) \over {z-\kappa}}\;.
\label{self}
\end{equation}

\noindent
$\tilde G^{-1}(z)$ has the Herglotz property only if $T(\kappa)$ is
non-negative.

In general, the integral in Eq. (\ref{self}) is divergent and therefore
needs renormalization. The usual mass and wave-function renormalizations
are performed by imposing the condition that the renormalized propagator has a
pole at the physical nucleon mass $M$ with unit residue. This implies
that the renormalized inverse propagator $\tilde G_{R}^{-1}(z)$, defined as

\begin{equation}
\tilde G_{R}^{-1}(z) \equiv Z_2 \tilde G^{-1}(z)\;,
\label{rgm}
\end{equation}

\noindent
is given by

\begin{equation}
\tilde G_{R}^{-1}(z) =(z-M)\left[ 1-(z-M)\int_{- \infty}^{+\infty} d\kappa
{T_{R}(\kappa) \over {(\kappa-M)^2(z-\kappa)}}\right]\;,
\label{rself}
\end{equation}

\noindent
where $T_{R}(\kappa)=Z_2 T(\kappa)$ and

\begin{equation}
Z_2=\left[ 1 - \int_{- \infty}^{+\infty} d\kappa {T_R(\kappa) \over
{(\kappa-M)^2}}\right]\;.
\label{z2}
\end{equation}

\noindent
{}From Eq. (\ref{rgm}), it follows
that the spectral representation for $\tilde G_R(z)$ is

\begin{equation}
\tilde G_R(z) = \int_{- \infty}^{+\infty} d\kappa {A_{R}(\kappa) \over
{z-\kappa}}\;,
\label{rg}
\end{equation}

\noindent
where $A_{R}(\kappa) = A(\kappa)/Z_2$. In terms of renormalized quantities,
$Z_2$ can be written as

\begin{eqnarray}
Z_2 &=& 1 - \int_{- \infty}^{+\infty} d\kappa {T_{R}(\kappa)
\over{(\kappa-M)^2}}\label{z2t}\\
   &=&
\left[\int_{- \infty}^{+\infty} d\kappa A_{R}(\kappa)\right]^{-1}\;.
\label{z2a}
\end{eqnarray}

In order to compare with previous work of BPW, we now consider a model with
nucleons ($\psi$), pions($\vec \pi$) and omegas($\omega^{\mu}$) given by the
following Lagrangian density,

\begin{eqnarray}
{\cal L} = &&{\bar \psi}(i \gamma_{\mu} \partial^{\mu}
-i g_{0\pi}\gamma_5 \vec \tau \cdot \vec \pi -
g_{0\omega} \gamma_{\mu} \omega^{\mu}) \psi \nonumber \\
&&- {1\over 4} F_{\mu \nu} F^{\mu \nu}
- {1\over 2} m_{\omega}^2 \omega_{\mu} \omega^{\mu}
+{1\over 2} \partial_{\mu} \vec \pi \cdot \partial^{\mu} \vec \pi
-{1\over 2} m_{\pi}^2 \vec \pi \cdot \vec \pi\;,
\label{lag}
\end{eqnarray}

\noindent
where
$F^{\mu \nu} = \partial^{\mu} \omega^{\nu} - \partial^{\nu} \omega^{\mu}$.
The Schwinger-Dyson equation for the nucleon propagator, Fig. {\ref{fig1}},
is given by

\begin{eqnarray}
G^{-1}(p) &=& G_0^{-1}(p)+3 i g_{0\pi}^2 \int {d^4q\over (2\pi)^4} \gamma_5
D_{\pi}(q^2)G(p-q)\Gamma_5(p-q,p;q) \nonumber \\
&+&i g_{0\omega}^2 \int {d^4q\over (2\pi)^4} \gamma_{\mu}
D_{\omega}^{\mu \nu}(q^2)G(p-q)\Gamma_{\nu}(p-q,p;q)\;,
\label{sde}
\end{eqnarray}

\noindent
where $D_{\pi}$ and $D^{\mu \nu}$ are the $\pi$ and $\omega$
propagators and $\Gamma_5(p-q,p;q)$ and $\Gamma_{\mu}(p-q,p;q)$ are the
pion-nucleon
and omega-nucleon vertex functions, respectively. We do not consider the
vector meson tadpole contribution to the nucleon propagator since it drops
out in the renormalization procedure. The Hartree-Fock (HF)
approximation amounts to use the non-interacting $D_{\pi}$ and
$D_{\omega}$ and the bare vertices $\Gamma_5(p-q,p;q) = \gamma_5$ and
$\Gamma_{\mu}(p-q,p;q)=\gamma_{\mu}$ in Eq. (\ref{sde}). In the HF
approximation
the set of equations to be solved self-consistently for the renormalized
$A_R(\kappa)$ and $T_R(\kappa)$ is given by

\begin{equation}
T_R(\kappa)=\int_{-\infty}^{+\infty} d\kappa'K(\kappa,\kappa')A_R(\kappa')
\label{tkap}
\end{equation}

\begin{equation}
\tilde G^{-1}_R(\kappa(1+i\epsilon))=(\kappa-M)\left[1-(\kappa-M)
\int_{-\infty}^{+\infty} d\kappa' {T_R(\kappa') \over
{ (M-\kappa')^2\left[\kappa'-\kappa(1+i\epsilon)\right]}}\right]
\label{set}
\end{equation}

\begin{equation}
A_R(\kappa)=\delta(M-\kappa) + |\tilde G^{-1}_R(\kappa(1+i\epsilon))|^{-2}
T_R(\kappa)\;,
\label{akap}
\end{equation}

\noindent
where $K(\kappa,\kappa')$ is given by

\begin{equation}
K(\kappa, \kappa') = K_{\pi}(\kappa, \kappa')+
K_{\omega}(\kappa, \kappa')\nonumber
\end{equation}

\begin{eqnarray}
K_{\pi}(\kappa, \kappa') &=& 3\left({g_{\pi}\over 4\pi}\right)^2
\left[\kappa^4-2\kappa^2({\kappa'}^2+m_{\pi}^2)+({\kappa'}^2
-m_{\pi}^2)^2\right]^{1/2}\nonumber \\
&\times&{1\over|\kappa|^3}\left[(\kappa-\kappa')^2-m_{\pi}^2\right]
\theta(\kappa^2-(|\kappa'|+m_{\pi})^2)
\label{kpi}
\end{eqnarray}

\begin{eqnarray}
K_{\omega}(\kappa, \kappa') &=& \left({g_{\omega}\over 4\pi}\right)^2
\left[\kappa^4-2\kappa^2({\kappa'}^2+m_{\omega}^2)+({\kappa'}^2-
m_{\omega}^2)^2\right]^{1/2}\nonumber \\
&\times&{1\over|\kappa|^3}\left[(\kappa-\kappa')^2-2\kappa \kappa'
-m_{\omega}^2\right]\theta(\kappa^2-(|\kappa'|+m_{\pi})^2)\;.
\label{kw}
\end{eqnarray}

\noindent
In the above equations, $g_{\pi}$ and $g_{\omega}$ are the renormalized
coupling constants, defined as $g_{\pi}=Z_2g_{0\pi}$ and
$g_{\omega}=Z_2g_{0\omega}$.

Following BPW, let us initially consider the pion, neglecting for the moment
the vector meson. Eqs. (\ref{tkap}-\ref{akap}) were solved numerically by
iteration, begining in Eq. (\ref{tkap}) with the free value for
$A(\kappa)$, $A(\kappa)=\delta(M-\kappa)$.  The converged function
$A(\kappa)$, for $g_{\pi}^2/4\pi=14.4$ and the physical nucleon and pion
masses, is shown by the dashed line in Fig. (\ref{fig2}).
In addition to the pole at the nucleon mass $z=M$
(which is fixed by the renormalization procedure), $\tilde G_R^{-1}(z)$ in
Eq. (\ref{rself}) has zeros at $z=(0.73\pm 1.25i)M$. These complex zeroes mean
that the nucleon propagator has poles at those complex masses with
coresponding residues of $-0.75\pm 0.32i$ respectively. The signal for the
presence of ghosts is revealed by the fact that $Z_2$ calculated from
$T_R(\kappa)$ in Eq. (\ref{z2t}) gives $Z_2=-\infty$. Since $Z_2^{-1}=0$, it
follows from Eq. (\ref{z2a}) that the integral of $A_R$ is zero. We must
therefore include the pair of complex conjugated poles in $\tilde G_R$:

\begin{equation}
\tilde G_R(z)= \int_{-\infty}^{\infty} d\kappa {A_R(\kappa)\over{z-\kappa} }
+ {A_c \over {z-\kappa_c}}+{A_c^* \over {z-\kappa_c^*}}\;,
\label{cres}
\end{equation}

\noindent
where $\kappa_c$ and $A_c$ are the complex pole and residue respectively. The
sum of residues of the complex poles is negative and exactly cancels the
integral over real $\kappa$. Actually, the difficulty lies in the negative
sign of $Z_2$, since this destroys the Herglotz property of $G^{-1}$
(unrenormalized).

The ghosts have their origin in the ultraviolet behavior of the interaction,
as we shall show in the following. The kernel $K_{\pi}(\kappa,\kappa')$ has the
following asymptotic form, for large $\kappa$ or $\kappa'$,

\begin{equation}
K_{\pi}(\kappa,\kappa') \longrightarrow {1\over 2|\kappa|^3}(\kappa^2-
\kappa'^2)(\kappa-\kappa')^2\theta(\kappa^2-\kappa'^2)\;.
\label{asyk}
\end{equation}

\noindent
Since $\int d\kappa A(\kappa)$ is finite\cite{bpw}, it follows from
Eqs. (\ref{tkap},\ref{asyk}) that $T(\kappa)$ for large $\kappa$ is
given by

\begin{equation}
T_R(\kappa) \longrightarrow |\kappa|\;.
\label{asyt}
\end{equation}

\noindent
Therefore, the integral in Eq. (\ref{z2t}) is divergent and
$Z_2 \longrightarrow -\infty$.

The $\omega$-meson introduces a new ingredient in the problem, namely the
spectral function $A_R(\kappa)$ can be negative for some values of real
$\kappa$.  The dotted line in Fig. (\ref{fig2}) represents $A_R(\kappa)$ for
$g_{\omega}^2/4\pi=6.36$ and $m_{\omega}=780\;MeV$. The complex poles are
located at $z=(5.67 \pm 11.76i)M$, with residues $-1.0412\pm 0.22i$
respectively. $A(\kappa)$ is negative for $M+m_{\omega} \leq \kappa < 3.9$.
The $K_{\omega}$ has a finite negative jump at $\kappa=M+m_{\omega}$ due
to the term $-2\kappa \kappa'$ in Eq. (\ref{kw}). This introduces a
discontinuity in the integrand of Eq. (\ref{set}) for $\tilde G_R^{-1}$.
At the discontinuity, the real part (principal value integral) in
Eq. (\ref{rself})  has a logarithmic sigularity, implying that $A_R(\kappa)$
has a (sharp) zero at $\kappa=M+m_{\omega}$. This zero is represented in
Fig. (\ref{fig2}) by the vertical straight line which hits the $\kappa$ axis
at the discontinuity .

Although negative $A_R(\kappa)$ represents presumably the
presence of negative metric states (sometimes also referred to as ghosts),
this is not related to the complex ghost poles which motivated this work. We
show below that the use of form factors will eliminate the complex poles,
but $A_R(\kappa)$ may still contain negative regions as before.
For QED, the interaction kernel is the same as in Eq. (\ref{kw}), with
$m_{\omega}=0$, giving rise to a negative spectral function as
well\cite{gelow}. In theories with massless vector bosons, the theory is
formulated in terms of an indefinite metric\cite{gubl}  and the positivity of
the spectral functions is not a necessary requirement\cite{gelow}.
However, in our case where $m_{\omega} \neq 0$, $A_R(\kappa) \geq 0$ is a
necessary requirement and the only explanation we have at the moment for
$A_R(\kappa) < 0$ is the inadequancy of the HF approximation, or to
the inconsistency of the theory.

Including both $\pi$- and $\omega$-mesons, the situation is qualitatively
similar as above. The contribution of the $\pi$ dominates the one of the
$\omega$ and $A_R(\kappa)$ is non-negative for real $\kappa$.
In Fig. (\ref{fig2}), the solid line indicates the function $A_R(\kappa)$ for
the same values of $g_{\pi}$, $m_{\pi}$, $g_{\omega}$ and $m_{\omega}$ as
above. The complex poles are located at $z=(1.05\mp 1.26i)M$ with residues
$-0.77\pm 0.20i$ respectively.

\section{VERTEX CORRECTIONS}

In this section we discuss the effect of form factors on the problem
of ghost poles. We start with the consideration of the Sudakov form factor.

Let us consider the proper $NN\omega$ vertex $\Gamma^{\mu}(p_1, p_2, q)$,
Fig. {\ref{fig3}(a)}. The $p_1$ and $p_2$ are the external nucleon
momenta and $q$ is the external meson momentum. The Sudakov form-factor is
obtained by summing the leading-log contributions of all vertex corrections.
For QED, for off-shell space-like nucleon momenta, the Sudakov form-factor
is given by

\begin{equation}
\Gamma^{\mu}(p_1, p_2, q) = \gamma^{\mu} \exp \left[{-{e^2\over 8\pi^2}
\ln\biggl\vert{q^2\over p_1^2}\biggr\vert\;\ln\biggl\vert{q^2\over p_2^2}
\biggr\vert}\right]\;.
\label{qedsud}
\end{equation}

\noindent
This expression is valid for large nucleon momenta, $\vert p_1^2\vert$,
$\vert p_2^2\vert \gg M^2$, and $\vert q^2\vert \gg \vert p_1^2\vert,
\vert p_2^2\vert $. Although the momenta appearing in Eq. (\ref{qedsud})
are space-like, Sullivan and Fishbane\cite{fi} argue that the expression may
be freely continued to the time-like region. In the case of massive vector
mesons, we have exactly the same expression, for large momenta, as in
Eq. (\ref{qedsud}), with $e^2$ replaced by $g_{\omega}^2$

Our approach consists in replacing the bare vertices by the corresponding
vector meson corrected ones. Of course, all values of the loop momentum $q^2$
are formally required in the evaluation of $\tilde G^{-1}(p)$ in
Eq. (\ref{sde}), corresponding to the exchanged mesons in Fig. (\ref{fig1}).
The lowest order (BPW) approximation, using the Hartree-Fock form of the
$\Gamma$'s, is correct only for $q=0$ and is not asymptotically correct for
high $q^2$ ($\Gamma_{\mu}$ and $\Gamma_5$ are entirely independent of $q$
in BPW). Since the Sudakov form, Eq. (\ref{qedsud}), introduces convergence
at high $q^2$, one might hope that our approach will eliminate the ghost
problem appearing in the ultraviolet. Although we will see that this is the
case, a fully satisfactory analysis requires knowing that the asymptotic high
$q^2$ behavior of the Sudakov form is correct.

Although the integral in Eq. (\ref{tkap}) requires all values of $q$, a
correct high $q$ behavior, together with a correct $q=0$ limit, should provide
at least a``ghost-free" result. Allendes and Serot calculated in first order
perturbation theory the (low $q^2$) on-shell vertex function and then
interpolated the result to the on-shell Sudakov form factor in the ultraviolet
(high $q^2$). Since the calculation of the low momentum behavior of the
off-shell vertex function is a tremendous task, even in lowest order
perturbation theory, we prefer not to follow the approach of Allendes and
Serot. We, instead, rewrite Eq. (\ref{qedsud}) as

\begin{equation}
\Gamma^{\mu}(p_1,p_2,q) = \gamma^{\mu} F(p_1,p_2,q)
\label{appr}
\end{equation}

\noindent
with

\begin{equation}
F(p_1,p_2,q)=\exp \left[{-{g_{\omega}^2\over 8\pi^2}
\ln\biggl\vert{{\Lambda^2+q^2} \over  {\Lambda^2+p_1^2}}\biggr\vert\;
\ln\biggl\vert{{\Lambda^2+q^2} \over {\Lambda^2+p_2^2}}\biggr\vert}\right]\;,
\label{f}
\end{equation}

\noindent
where $\Lambda$ is an infrared cut-off which will be fixed later.
The effect of $\Lambda$ is to extend the validity of form factor to the
infrared region of the loop integral. The important momentum dependence of
the form factor for the elimination of the ghosts is the $p_1$ (or $p_2$)
dependence. We note that the region of large $q^2$ does not contribute in the
loop integral, and so the conditions for the validity of the form factor in
the ultraviolet region of the loop integral are satisfied in practice.

It is not difficult to show, following Sudakov's derivation\cite{sud}, that
the vector meson correction to the proper $NN\pi$ vertex,
Fig. {\ref{fig3}(b)},
gives the same exponential suppression as given by Eq. (\ref{qedsud}). We write
the $NN\pi$ vertex function as

\begin{equation}
\Gamma_5(p_1, p_2, q) = \gamma_5 F(p_1,p_2,q)
\label{pisud}
\end{equation}

\noindent
where we have also introduced an infrared regulator. With the introduction of
the infrared regulator, the vertex functions have the correct zero momentum
limit $F(0,0,0)=1.$

By substituting the Sudakov corrected vertex functions in the
Schwinger-Dyson equation Eq. (\ref{sde}), we obtain the following expression
for the Sudakov corrected kernels

\begin{equation}
K_{\pi}^{Sud}(\kappa, \kappa') = K_{\pi}(\kappa,\kappa')\;
F(\kappa,\kappa',m_{\pi})
\label{kerpi}
\end{equation}

\noindent
and

\begin{equation}
K_{\omega}^{Sud}(\kappa, \kappa') = K_{\omega}(\kappa,\kappa')\;
F(\kappa,\kappa',m_{\omega})
\label{kerom}
\end{equation}

\noindent
where $K_{\pi}(\kappa, \kappa')$ and $K_{\omega}(\kappa, \kappa')$ are the
``bare" interaction kernels of Eqs. ({\ref{kpi}}-{\ref{kw}}).

We have also studied the ghost problem in the context of purely parametrized
form factors, similar to the ones employed in boson exchange nucleon-nucleon
potential models\cite{bonn}. There is one complication in our case, namely we
need off-shell form factors, since the external nucleon legs are off-shell.
Here we follow Ref. (\cite{gros}) and use one supression factor for each
external leg of the vertex

\begin{equation}
F(p_1,p_2,q)={1\over{1+|p_1^2/\Lambda^2|}}\;{1\over{1+|q^2/\Lambda^2|}}\;
{1\over{1+|p_2^2/\Lambda^2|}}\;,
\label{parf}
\end{equation}

\noindent
where the value of $\Lambda$ is discussed in the next section.

Substitution of this is the Schwinger-Dyson-Equation Eq. (\ref{sde}), we
obtain Eqs. (\ref{kerpi}-\ref{kerom}) for the corrected kernels.

\section{NUMERICAL RESULTS}

We start discussing the Sudakov supression for the pion. The iterative solution
of the equation proceeds as before. We find that the ghost poles disappear for
values of $\Lambda$ smaller than a critical value
$\Lambda_{crit} \approx 0.9\;M$. The function $A_R(\kappa)$ is shown by the
dashed line in Fig. (\ref{fig4}) for $\Lambda=\Lambda_{crit}$. Besides the
fact that we do not find complex poles, we also find that

\begin{equation}
Z_2 = 1 - \int_{- \infty}^{+\infty} d\kappa {T_{R}(\kappa)
\over{(\kappa-M)^2}}=\left[\int_{- \infty}^{+\infty} d\kappa
A_{R}(\kappa)\right]^{-1}=0.16\;.
\label{sz2pi}
\end{equation}

\noindent
The important point about Eq. (\ref{sz2pi}) is that we get a positive
$Z_2$, and so we do not have the usual sign for the presence of ghosts.
Moreover, we obtain the same value for $Z_2$ calculated either with
$A_R(\kappa)$ or with $T_R(\kappa)$, meaning that we do not have missing
strength and so, $\tilde G_R^{-1}(z)$ given by Eq. (\ref{rself}) is the
inverse of $G_R$ given by Eq. (\ref{rg})\cite{bpw}.

For the $\omega$, we find that for a $\Lambda_{crit}\approx 0.55\;M$, the
ghosts disappear. The dashed line in Fig. (\ref{fig4}) represents
the function $A_R(\kappa)$ for $\Lambda=0.55\;M$. Fig. (\ref{fig4}) shows that,
as mentioned in Section II, although the ghosts poles have disappeared,
$A_R(\kappa)$ contains a negative region as before. The obvious conclusion is
that the negative metric states {\it are not} related to the BPW ghost poles.

Including both $\pi$ and $\omega$ mesons, the situation is qualitatively
similar as for the $\pi$ and $\omega$ in isolation. The solid curve in
Fig. (\ref{fig4}) refers to a $\Lambda=0.6\;M$ for the $\pi$ and
$\Lambda=0.55\;M$. The different values for $\Lambda_{crit}$ for $\pi$ and
$\omega$ is understood on the basis of their range in the $NN$ force. This is
clearly manifest in combined the case of $\pi$ and $\omega$, where the
supression is governed entirely by a $\Lambda$ closely to the
$\Lambda_{crit}$ for the $\omega$.

In Fig. (\ref{fig5}) we present the results for the spectral function when
using the monopole corrected vertices of Eq. (\ref{parf}). The situation is
similar to the Sudakov supression with respect to the role of the cutoff and
the disappearance of the ghost poles. For $\Lambda$'s smaller than a critical
value of $\Lambda$, the complex poles disappear and $Z_2$ calculated via $A_R$
and $T_R$ give identical results. For the plots in Fig. (\ref{fig5}) we have
used $\Lambda=M$ for all cases. In Fig. (\ref{fig6}) we illustrate the
trajectory of the complex poles as a function of $\Lambda$, for the case of
the $\pi$. The critical value of $\Lambda$ for the disappearance of the
ghosts is of the same order (Fig. (\ref{fig6})) as used in potential
models\cite{bonn}. One interesting feature of the different form factors is
that the Sudakov form factor ``squeezes" the spectral function around
$\kappa=M+m_{\omega}$, while the monopole form factor has the effect
of decreasing the spectral function all over the $\kappa$ range. This is
possibly due to the fact that the Sudakov form factor provides a strong
supression for high momenta only, while the monopole form factors provide
an uniform supression starting at momenta of the order of the nucleon
mass.

\section{CONCLUSIONS}
The nucleon propagator treated in the Hartree-Fock approximation in a
hadronic field theory is plagued with the presence of ghost poles. The
appearance of the ghosts is related to the ultraviolet behavior of the
nucleon-meson interaction. In a theory containing vector mesons, the
infrared structure of such a theory introduces corrections to the various
nucleon-meson vertex functions which are strongly damped in the ultraviolet.
The damping arises from the dressing of these vertices by the vector mesons.
In this work we have shown that such a damping of the vertex function can
eliminate the appearance of the ghost poles of the usual Hartree-Fock
approximation.

We emphasize that our work has to be extended and improved on several aspects
before definite conclusions can be drawn for realistic calculations in nuclear
physics. First, although the ghost problem has been shown not to be related to
the presence of a Fermi sea, the effects of the Sudakov form factor in this
case is worth to be investigated. In our work we have simply used the Sudakov
form factor in the loop integral of the Schwinger-Dyson equation. The
off-shell behavior of the vertex function at low $q^2$ has to be investigated.

In summary, we have shown that the inclusion of corrections due to the
dressing of vector mesons of nucleon-meson vertex functions are able to soften
the ultraviolet behavior of hadronic field theories. Therefore, vacuum
corrections to mean field approximations can be calculated in a consistent way.

\newpage
\figure{Diagrammatic representation of the Schwinger-Dyson equation
for the full nucleon propagator. The wavy (dashed) line represents the
$\omega$ ($\pi$) meson propagator, and the solid (double solid) line
represents the free (full) nucleon propagator.
\label{fig1}}
\figure{Spectral function $A_R(\kappa)$ for $\pi$ (dashed),
$\omega$ (dotted) and $\pi + \omega$ (solid).
\label{fig2}}
\figure{Diagrammatic representation of vertex corrections for the
(a) $NN\omega$ vertex and (b)$NN\pi$ vertex.
\label{fig3}}
\figure{Spectral function $A_R(\kappa)$ including the Sudakov form factor.
The meaning of the different lines is the same as in Fig. (\ref{fig2}).
For values of $\Lambda$ see text.
\label{fig4}}
\figure{Spectral function $A_R(\kappa)$ including the monopole form factors.
The meaning of the different lines is the same as in Fig. (\ref{fig4}).
For values of $\Lambda$ see text.
\label{fig5}}
\figure{Trajectory of the real (solid) and imaginary (dashed) parts of the
upper complex pole of the nucleon propagator for the $\pi$ contribution as a
function of the cutoff parameter in the monopole form factors. The arrow
indicates the critical value of $\Lambda$ for the disappearance of the ghost
poles.
\label{fig6}}
\end{narrowtext}
\end{document}